\documentclass[aps,prb,twocolumn,10pt,longbibliography,superscriptaddress,floatfix]{revtex4-1}
\usepackage{epsfig,amsmath,amssymb,color,comment}
\usepackage[bookmarks=true,colorlinks,linkcolor=blue,urlcolor=blue,citecolor=blue]{hyperref}

\usepackage{bm}
\usepackage{dsfont}%added by PJU

\definecolor{michael}{rgb}{0,.8,.5}

\newcommand\RR{{\mathds{R}}}

\begin{document}

\title{Prethermalization and Persistent Order in the Absence of a Thermal Phase Transition}

\author{Jad C. Halimeh}
\affiliation{Physics Department and Arnold Sommerfeld Center for Theoretical Physics, Ludwig-Maximilians-Universit\"at M\"unchen, D-80333 M\"unchen, Germany}

\author{Valentin Zauner-Stauber}
\affiliation{Vienna Center for Quantum Technology, University of Vienna, Boltzmanngasse 5, 1090 Wien, Austria}

\author{Ian P. McCulloch}
\affiliation{ARC Centre for Engineered Quantum Systems, School of Mathematics and Physics,
The University of Queensland, St Lucia, Queensland 4072, Australia}

\author{In\'{e}s de Vega}
\affiliation{Physics Department and Arnold Sommerfeld Center for Theoretical Physics, Ludwig-Maximilians-Universit\"at M\"unchen, D-80333 M\"unchen, Germany}

\author{Ulrich Schollw\"{o}ck}
\affiliation{Physics Department and Arnold Sommerfeld Center for Theoretical Physics, Ludwig-Maximilians-Universit\"at M\"unchen, D-80333 M\"unchen, Germany}

\author{Michael Kastner}
\affiliation{National Institute for Theoretical Physics, Stellenbosch 7600, South Africa}
\affiliation{Institute of Theoretical Physics, Department of Physics, University of Stellenbosch, Stellenbosch 7600, South Africa}

\date{\today}

\begin{abstract}
We numerically study the dynamics after a parameter quench in the one-dimensional transverse-field Ising model with long-range interactions ($\propto 1/r^\alpha$ with distance $r$), for finite chains and also directly in the thermodynamic limit. In nonequilibrium, i.e., before the system settles into a thermal state, we find a long-lived regime that is characterized by a prethermal value of the magnetization, which in general differs from its thermal value. We find that the ferromagnetic phase is stabilized dynamically: as a function of the quench parameter, the prethermal magnetization shows a transition between a symmetry-broken and a symmetric phase, even for those values of $\alpha$ for which no finite-temperature transition occurs in equilibrium. The dynamical critical point is shifted with respect to the equilibrium one, and the shift is found to depend on $\alpha$ as well as on the quench parameters. 
\end{abstract}
\maketitle

\section{Introduction}

In equilibrium, phase transitions and critical phenomena are well established and much studied, and implications like universality and scaling are well understood. Extending these concepts to nonequilibrium is a topic of active research. Several fundamentally different notions of so-called dynamical phase transitions have been proposed, but their mutual relations, and also the associated universality classes and scaling laws, are only poorly understood. In this paper we are concerned with a type of dynamical phase transition that is based on the notion of an order parameter, similar to Landau's theory of phase transitions in equilibrium. The key idea is to identify a dynamical phase transition on the basis of a suitable order parameter in a prethermal regime,\cite{EcksteinKollar08,EcksteinKollarWerner09,TsujiEcksteinWerner13} i.e., a nonequilibrium regime in which the system may be found before relaxing to thermal equilibrium, and which persists sufficiently long such that a value can be assigned to the order parameter.\cite{MoeckelKehrein08,MoeckelKehrein10,SciollaBiroli10,SciollaBiroli11,GambassiCalabrese11,SciollaBiroli13,Chandran_etal13,Maraga_etal15,Smacchia_etal15} A prethermal state retains some memory of the initial state of the system, therefore the prethermal value of the order parameter will in general differ from its thermal equilibrium value, and it may or may not show symmetry breaking and other signatures associated with the occurrence of a phase transition.

A simple protocol for probing such a dynamical phase transition is a quantum quench into the vicinity of an equilibrium quantum critical point. Consider a family of Hamiltonians $H(\lambda)=H_1 + \lambda H_2$, parametrized by $\lambda\in\RR$. In equilibrium at zero temperature and some critical parameter value $\lambda_\text{c}$, a quantum phase transition will in many cases occur, i.e., an abrupt change of the ground state properties of $H$. The idea of a quantum quench is to prepare the system in the ground state of $H(\lambda_0)$, and then, starting at time $t=0$, time-evolve that state under $H(\lambda)$ with $\lambda\neq\lambda_0$. Depending on the quench parameters and the system under investigation, signatures similar to those of the equilibrium phase transition may or may not persist and be visible after the quench, critical properties may be modified, enhanced, or attenuated. Questions of this sort have previously been addressed mostly in mean-field models\cite{SciollaBiroli10,SciollaBiroli11} and field theories.\cite{GambassiCalabrese11,SciollaBiroli13,Chandran_etal13}

Dynamical phase transitions are expected to be related in some way to their equilibrium counterparts, as they show a similar kind of symmetry-breaking and are signalled by the same order parameter.  Whether such a relation exists in all cases, and what its precise nature is, is a question that we want to address in this paper. A relation to equilibrium quantum phase transitions at $T=0$ is supported by the fact that in previous work dynamical phase transitions have been observed by quenching into the vicinity of a quantum critical point. Additionally, a relation to a finite-$T$ phase transition may be conjectured by noticing that a quench populates excited states above the ground state of the postquench Hamiltonian, which generically, at least after sufficiently long times, are expected to approach a thermal distribution with $T>0$.

\begin{figure}\centering
\includegraphics[width=0.95 \columnwidth]{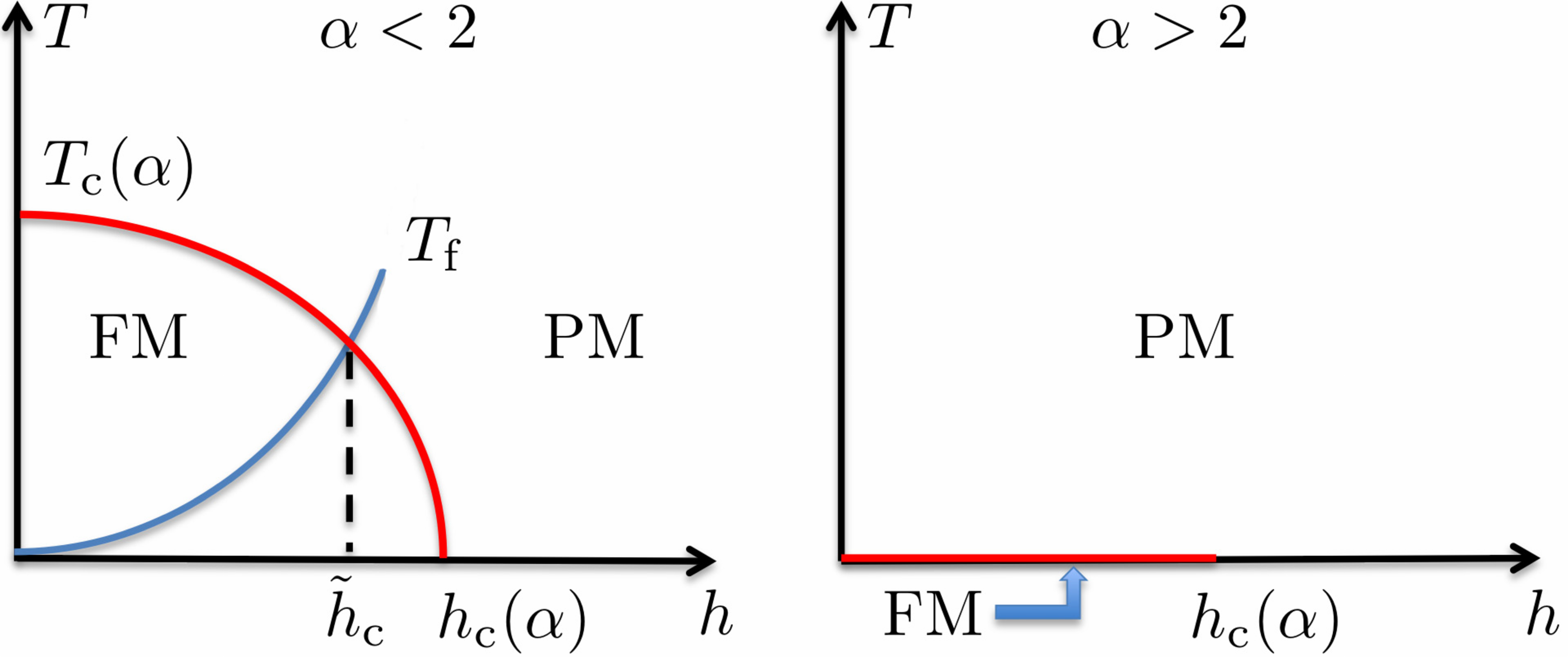}
\caption{\label{fig:illustrative}
Schematic phase diagram of the long-range TFIM \eqref{eq:TFIM}. The model exhibits an equilibrium quantum phase transition at a critical point $h_c(\alpha)$ for all values of $\alpha$. A finite-$T$ phase transition occurs only for $\alpha<2$ (left), but not for $\alpha>2$ (right). Quenching from $h_\text{i}=0$ to $h_\text{f}=h$ and letting the system thermalize, equilibrium states on a line $T_\text{f}(h)$ (blue line in the left plot) are reached at long times. The phase transition line is crossed at a field $\tilde{h}_\text{c}<h_\text{c}(\alpha)$, which results in a shift of the critical field with respect to the quantum critical point.
}
\end{figure}

\section{Long-range transverse-field Ising model}

To probe the relation between equilibrium and dynamical phase transitions, we study a model that has a quantum phase transition at zero temperature, and additionally, depending on a parameter, may or may not have a finite-$T$ transition as well. A model that has these desired properties is the transverse-field Ising model (TFIM) with power-law interactions,
\begin{equation}\label{eq:TFIM}
H(h) = -J\sum_{i>j=1}^L\frac{\sigma^z_i\sigma^z_j}{|i-j|^{\alpha}}-h\sum_{i=1}^L\sigma^x_i
\end{equation}
where $J>0$ is a spin--spin coupling constant. We consider one-dimensional lattices consisting of $L$ lattice sites, and $\sigma_i^a$ with $a\in\{x,y,z\}$ denote the components of Pauli spin-$1/2$ operators on lattice site $i$. The exponent $\alpha$ in \eqref{eq:TFIM} tunes the range of the spin--spin interaction, from all-to-all coupling at $\alpha=0$ to nearest-neighbor coupling in the limit $\alpha\to\infty$. We restrict the discussion to exponents $\alpha>1$, so that an $N$-dependent scaling factor to make the Hamiltonian \eqref{eq:TFIM} extensive is not needed. For all values of $\alpha$, this model has a quantum phase transition at some critical magnetic field $h_\text{c}(\alpha)$, whereas a finite-$T$ phase transition occurs only for $\alpha\leq2$ (see Fig.~\ref{fig:illustrative} for an illustration).\cite{Dyson69a,DuttaBhattacharjee01}

We use the magnetic field $h$ as a quench parameter, starting in the groundstate $|\psi_\text{i}\rangle$ of an initial Hamiltonian $H(h_\text{i})$ at time $t=0$, and then time-evolving that state under the evolution generated by a Hamiltonian $H(h_\text{f})$ with a field $h_\text{f}$ different from $h_\text{i}$. We will mainly consider quenches starting from $h_\text{i}=0$, i.e., initial states from the degenerate ground space, where we pick the symmetry-broken, fully polarized state in $+z$ direction. Our aim is to detect the occurrence of a dynamical phase transition by monitoring the magnetization
\begin{equation}\label{e:m}
m(t)=\frac{1}{L}\sum_{j=1}^L \langle\psi_\text{i}(t)|\sigma_j^z|\psi_\text{i}(t)\rangle,
\end{equation}
where $|\psi_\text{i}(t)\rangle=\exp(-\text{i}H(h_\text{f})t)|\psi_\text{i}\rangle$ is the time-evolved state after the quench.

Except for the extreme cases $\alpha=0$ and $\alpha=\infty$, the model \eqref{eq:TFIM} is nonintegrable, and is expected to thermalize in the long-time limit. Hence, in that limit, the magnetization \eqref{e:m} will show order-parameter-like behavior for $\alpha<2$, or be vanishing throughout for $\alpha>2$, as predicted by the phase diagrams in Fig.~\ref{fig:illustrative}. While thermalization will happen eventually, the corresponding timescale can be extremely long, so long in fact that it may become irrelevant for experimental observations.

\section{Dynamical phase transitions}

A dynamical phase transition may be detected by studying the order parameter $m$ as a function of the final quench parameter $h_\text{f}$ in a nonequilibrium regime corresponding to intermediate timescales. To generate some intuition on what kind of behavior to expect, it is instructive to consider two limiting cases: (i) For small quenches from $h_\text{i}=0$ to $h_\text{f}\gtrsim0$, excitations above the groundstate of $H(h_\text{f})$ are only sparsely populated, the dynamics towards a finite-$T$ thermal state of Hamiltonian will take place very slowly, and a memory of the nonvanishing magnetization of the groundstate of $H(0)$ will be retained for a long time. (ii) For a large quench well beyond the critical point, $h_\text{f}\gg h_\text{c}(\alpha)$, excitations are massively populated, no slow variables are expected to exist, and a rapid approach to $m=0$ is expected. In between these two extreme cases (i) and (ii), one may expect a transition between a regime with nonvanishing magnetization at small $h_\text{f}$ and a regime with vanishing $m$ at large $h_\text{f}$. Such a dynamical phase transition has previously been observed in the TFIM with all-to-all interactions ($\alpha=0$),\cite{Das_etal06,Smacchia_etal15,ZunkovicSilvaFabrizio16} but this case is special in more than one way and its behavior is not expected to be generic.

\begin{figure}\centering
\includegraphics[width=0.9\linewidth]{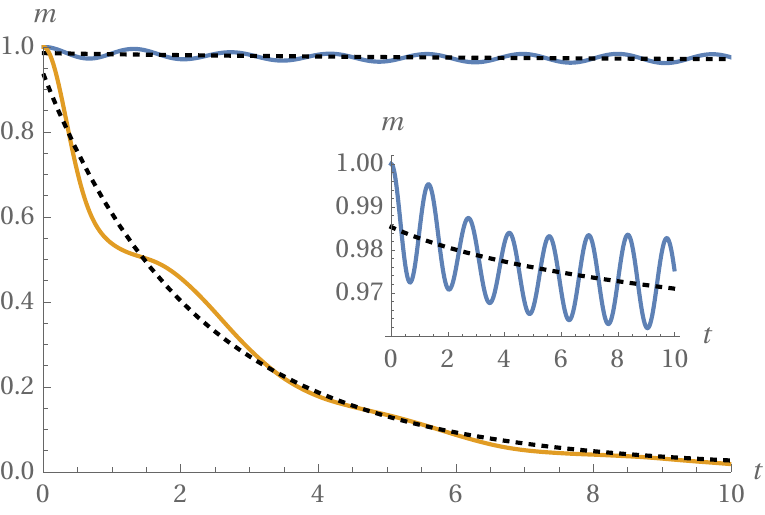}
\caption{\label{f:mvstFits}%
Time-evolution of the order parameter $m$ as obtained from iMPS simulations for long-range exponent $\alpha=3$. For a strong quench from $h_\text{i}=0$ to $h_\text{f}=0.99$, the magnetization quickly decays towards zero (yellow line) and is well approximated by a power law (lower black line). For a small quench from $h_\text{i}=0$ to $h_\text{f}=0.28$, the magnetization shows an initial decay away from its initial value of $1$ on a fast timescale (inset), and then saturates to a nonzero value for rather long times (blue line). Eventually, for the chosen parameter values and on a timescale not accessible in simulations, the system will thermalize to a state with zero magnetization.  
}%
\end{figure}

\section{Numerical methods}

In this paper we use two complementary numerical methods to study dynamical quantum phase transitions after a quench in the general (nonintegrable) TFIM with long-range interactions \eqref{eq:TFIM}. The first is the time-dependent density matrix renormalization group ($t$-DMRG) \cite{White92,Schollwoeck05,Schollwoeck11,WhiteFeiguin04,Verstraete_etal04,Vidal04,Daley_etal04,Gobert_etal05}  method with Krylov\cite{Krylov31} time evolution, which we apply to finite chains of up to 128 sites. The second is a novel method, based on a time-dependent variational principle for matrix product states,\cite{Schollwoeck11,VerstraeteMurgCirac08,Haegeman_etal11,Haegeman_etal} tailored for simulating the dynamics of long-range lattice systems in the thermodynamic limit. Details on this numerical method, which we abbreviate by iMPS, are provided in the companion paper \onlinecite{Halimeh_etal}. The combination of the two methods allows us to observe finite-size effects as would be visible in experimental realization on the one side, but also clean infinite-system idealizations as they are used in theoretical approaches. Both numerical methods are certified in the sense that they use well controlled approximations, tunable by an upper bound of the entanglement of the simulated states, which we set to achieve good simulation accuracies. During the simulation we monitor the order parameter $m$ as a function of time \eqref{e:m}, as illustrated in Fig.~\ref{f:mvstFits} for different quench-parameters. The timescales that can be reached in the simulation depend on the lattice size $L$, but also on other system and quench parameters. The simulation methods used are considered the current state of the art for one-dimensional spin systems.

\section{Thermal behavior after a quench}

Before discussing dynamical phase transitions at intermediate times, it is instructive to discuss the thermal state reached at very long times after a quench. Starting in the groundstate corresponding to $h_\text{i}=0$ and quenching to $h_\text{f}\neq 0$, the system will not be in the groundstate of $H(h_\text{f})$. A nonintegrable model like the one we are studying is then believed to thermalize after a sufficiently long time towards a finite-temperature Gibbs state. The temperature of that state depends on $h_\text{f}$, and this dependence can be described by some function $T_\text{f}(h_\text{f})$. This implies that, by performing a quench and waiting sufficiently long for the system to thermalize, one explores the $(T,h)$ equilibrium phase diagram along the line $(T_\text{f}(h_\text{f}),h_\text{f})$ parametrized by $h_\text{f}$ (blue line in Fig.~\ref{fig:illustrative}). A phase transition will be observed for all $\alpha\leq2$ as predicted by equilibrium thermodynamics, and it will occur at a critical field $\tilde{h}_\text{c}$ (corresponding to the value at which $T_\text{f}(h)$ crosses the thermal equilibrium transition line) that is smaller than $h_\text{c}$ of the quantum phase transition. 

\begin{figure}[t]\centering
\includegraphics[width=0.9\linewidth]{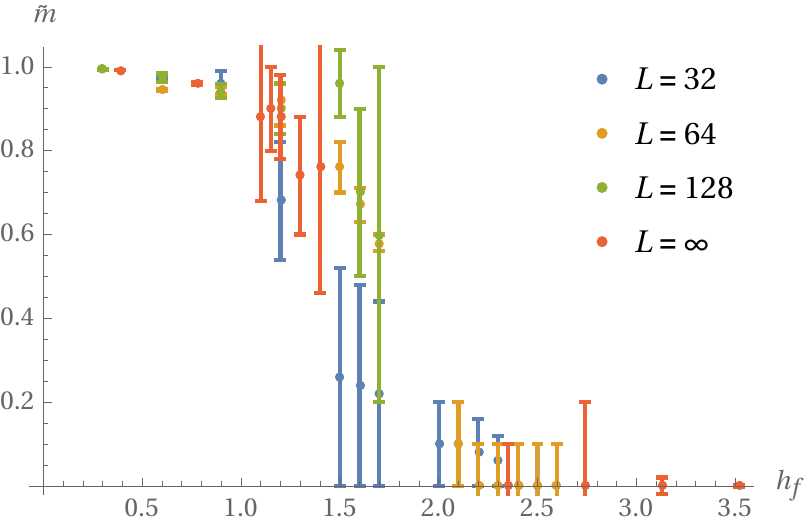}
\includegraphics[width=0.9\linewidth]{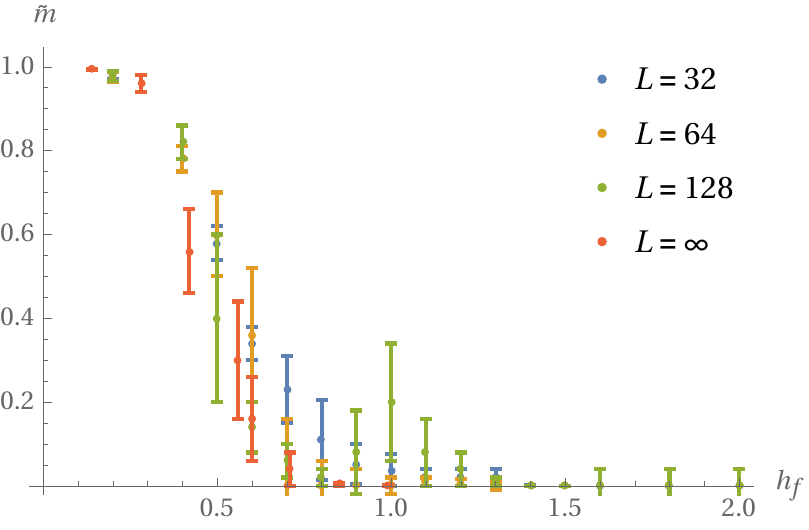}
\caption{\label{f:mofh}%
Prethermal magnetization $\tilde{m}$ plotted as a function of the final quench parameter $h_\text{f}$. Both plots are for quenches starting from $h_\text{i}=0$, and for various system sizes as indicated in the legends. The existence of a magnetized phase for small $h_\text{f}$ and an unmagnetized phase for large $h_\text{f}$ is clearly visible for $\alpha=1.6$ (top) and $\alpha=3$ (bottom).  
}%
\end{figure}

\section{Dynamical phase transition of the long-range TFIM}

Quenching and waiting for thermalization to occur is therefore not a way of observing novel nonequilibrium physics. To probe dynamical features we have to look at shorter timescales. The inset of Fig.~\ref{f:mvstFits} indicates that it is indeed reasonable and beneficial to use equilibrium concepts for the description of nonequilibrium observations on intermediate timescales. The magnetisation in that plot starts at $1$, and quickly decays away from this value to reach a plateau of $0.97$ around which it oscillates for the times reached in simulations. This prethermal value differs from thermal equilibrium, which is known to be $m=0$ for the parameters used in Fig.~\ref{f:mvstFits}. Our aim is to extract from the simulation data such prethermal magnetization values, which are indicative of the nonequilibrium physics on intermediate timescales relevant in various experimental settings. For some parameter values, the quasi-stationary value $\tilde{m}$ of the prethermal magnetization is clearly visible and easy to extract, while in other cases the limited simulation times compromise the accuracy and lead to large errorbars in the extracted values (see Appendix \ref{a:fitting} for details on the fitting procedure). 

Plotting the thus obtained prethermal magnetization $\tilde{m}$ as a function of the quench parameter $h_\text{f}$, we find a behavior that is reminiscent of an order parameter; see Fig.~\ref{f:mofh}. Due to the errorbars of $\tilde{m}$ it is difficult to determine the precise transition point of this dynamical phase transition on the basis of our numerical data, but we can confirm the existence of a magnetized phase for small quenches, and an unmagnetized phase for large quenches. Remarkably, the magnetized phase is clearly visible also for $\alpha=3$, and hence the dynamical phase diagram in this case differs drastically from its equilibrium counterpart, which does not have a ferromagnetic phase for $\alpha>2$. The comparison with iMPS data for infinite lattices confirms that this finding is not a finite-size artefact and indeed persists in the thermodynamic limit. Unfortunately, the (rather conservatively estimated) error bars in Fig.~\ref{f:mofh} do not allow to clearly establish whether or not the transition from the magnetized to the unmagnetized phase is indeed a sharp one, or to even extract critical exponents of such a dynamical phase transition. As is evident from Fig.~\ref{f:mofh}, the critical field $\tilde{h}_\text{c}$ at which the transition occurs becomes smaller for larger exponents $\alpha$. This suggests that for such shorter-ranged interactions the prethermalized state can be dynamically stabilized only for smaller quenches, and in that sense the ferromagnetically ordered state is less robust. We expect that $\tilde{h}_\text{c}$ approaches $h_\text{i}$ in the limit $\alpha\to\infty$, in agreement with the observation that exponential decay to the (generalized Gibbs) equilibrium value sets in immediately in the TFIM with nearest neighbor interactions.

\begin{figure}[t]\centering
\includegraphics[width=0.9\linewidth]{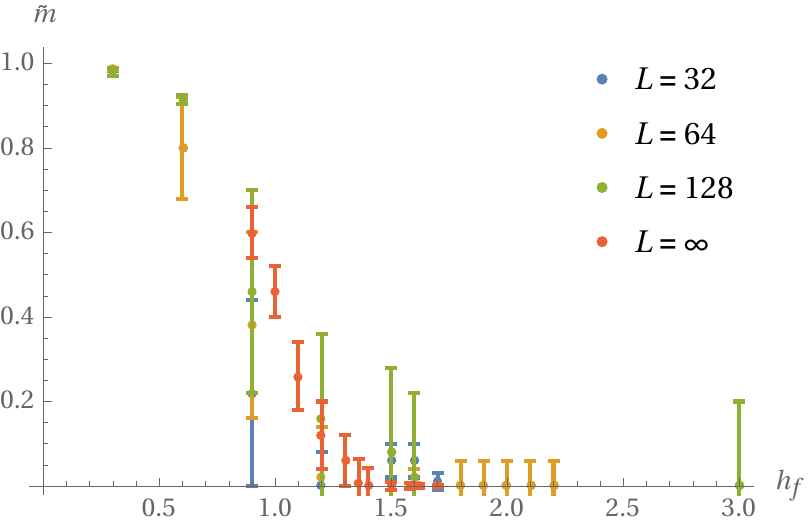}
\includegraphics[width=0.9\linewidth]{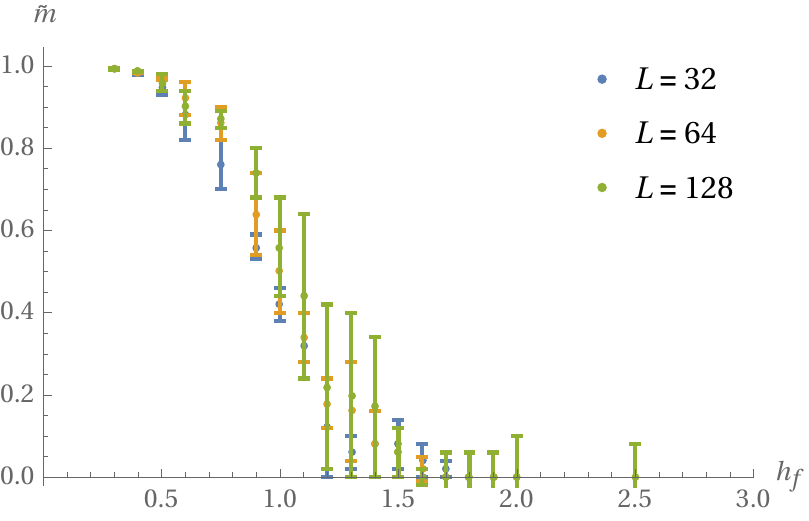}
\caption{\label{f:mofh2}%
Prethermal magnetization $\tilde{m}$ plotted as a function of the final quench parameter $h_\text{f}$ for $\alpha=2.3$. Top: quenching from $h_\text{i}=0$; bottom: quenching from $h_\text{i}=0.2$. Both plots show qualitatively similar behavior. A slight dependence of the dynamical critical point on $h_\text{i}$, as expected for the thermal behavior in the long-time limit after the quench, might also be present in the prethermalized regime on intermediate timescales, but cannot be established beyond doubt. 
}%
\end{figure}

\section{Other types of quenches}

As is usually the case in critical phenomena, the dynamical critical point is expected to be nonuniversal, but to depend on details of the Hamiltonian and, in our case, also on the quench protocol, in particular the initial quench parameter $h_\text{i}$. From the above discussion of the thermal equilibrium behavior after a quench, it appears plausible that also the dynamical critical point $\tilde{h}_\text{c}$ should be shifted towards slightly larger values with increasing $h_\text{i}$. To probe this effect, we consider quenches with different prequench Hamiltonians $H(h_\text{i})$, using initial fields $h_i=0$ and $0.2$. In Fig.~\ref{f:mofh2} we show and compare the corresponding dynamical phase diagrams. In both cases the transition from a dynamically ordered to a disordered phase is clearly established, but a shift of the dynamical transition point, if present, is concealed by numerical noise. 

It would be interesting to complement the results presented in this paper by studying quenches in the opposite direction, i.e., starting from the fully $x$-polarized groundstate of the Hamiltonian \eqref{eq:TFIM} in the limit $h_\text{i}\to\infty$ and quenching towards and across the quantum critical point from above. This setting is somewhat more difficult to investigate numerically, as in this case,
in addition to the Hamiltonian, the initial state is also $\mathbb{Z}_2$ symmetric. As a consequence, the magnetization is zero for all times and cannot be used to detect a dynamical phase transition. Alternatively, one could use second cumulants of the order parameter as done in Ref.~\onlinecite{Smacchia_etal15}, but such a signal is difficult to detect on the basis of limited-time data. Another possibility is to detect critical behavior on the basis of a diverging correlation length, as proposed in Ref.~\onlinecite{Nicklas_etal15}, but such an approach is tricky in long-range models, where, even away from criticality, ground state correlations are in general not exponentially clustered and hence the correlation length is diverging (or ill-defined).

\section{Conclusions}

In summary, we have studied the occurrence of a dynamical phase transition after a quench of the magnetic field in a transverse-field Ising model with long-range interactions. We have provided evidence that a symmetry-broken, ferromagnetic phase can be stabilized dynamically, in the sense that it persists for intermediate times in a prethermalized regime, even in the absence of a ferromagnetically-ordered equilibrium phase at finite temperature. We studied the dependence of such a dynamical phase transition on model parameters and quench parameters, in particular on the long-range exponent $\alpha$ and the pre-quench magnetic field $h_\text{i}$. While a specific model was chosen for the numerical study, we expect our findings to be valid more generally for long-range models, also in higher lattice dimension. 

The question studied in this paper is a numerically challenging one, and our results are obtained by state-of-the-art implementations of $t$-DMRG for finite one-dimensional lattices and an iMPS variational principle for infinite lattices. The latter is a novel approach, particularly suited for the problem at hand. An experimental investigation of the phenomena described in this paper should also be feasible: one-\cite{Richerme_etal14,Jurcevic_etal14} or two-dimensional\cite{Britton_etal12} arrays of trapped ions allow for the emulation of long-range interacting Ising spins in a magnetic field and, at least in principle, long-range exponents can be tuned in the range $0\leq\alpha\leq3$.\cite{PorrasCirac04} The required timescales, like in the numerical simulations, are an issue, but do not seem entirely out of reach. 

When finishing up this work we became aware of a preprint by Zunkovic {\em et al.}\cite{Zunkovic_etal} that addresses a similar question, but reaches different conclusions. In particular, the finite-size scaling extrapolations of Ref.~\onlinecite{Zunkovic_etal} are inconsistent with our infinite-system data.
\begin{acknowledgments}
The authors would like to thank Pasquale Calabrese, Damian Draxler, Martin Eckstein, Jutho Haegeman, Stefan Kehrein, Francesco Piazza, and Frank Verstraete for fruitful discussions.
V.\,Z.-S.\ acknowledges financial support by the Austrian Science Fund (FWF), grants F4104 SFB ViCoM and F4014 SFB FoQuS.
I.\,P.\,M.\ acknowledges support from the
Australian Research Council (ARC) Centre of Excellence for Engineered Quantum Systems, grant CE110001013 and the ARC Future Fellowships scheme, FT140100625.
I.\,d.\,V.\ acknowledges support by the Nanosystems Initiative Munich (NIM) (project No. 862050-2).
M.\,K.\ acknowledges support from the National Research Foundation of South Africa via the Incentive Funding and the Competitive Programme for Rated Researchers.  
\end{acknowledgments}

\appendix

\section{Fitting procedure}
\label{a:fitting}

When fitting numerical data like those in Fig.~\ref{f:mvstFits}, the situation can be summarized as follows: We have data of high accuracy, limited to an interval of times up to the order of 10 [setting the coupling constant $J=1$ in \eqref{eq:TFIM}]. The data typically show a decaying tendency, with fairly strong oscillations superimposed, like in the inset of Fig.~\ref{f:mvstFits}. Our aim is to extrapolate the decay to intermediate times that are, say, an order of magnitude longer than the times reached in the simulations. This timescale of extrapolation is reasonable for several reasons: (i) It is substantially longer than the timescale on which the decay to a prethermalization plateau occurs, hence we look at a timescale that is well separated from the initial dephasing dynamics. (ii) It is at least comparable to the timescales that, with some optimism, might be reached in experimental implementations. (iii) The timescale is short enough such that the errorbar that propagates to the extrapolated value is manageable. (Extrapolations to times that are orders of magnitude longer than the simulated times simply become unreliable.)

The main difficulty arises from the fact that the functional form of the decay is not known. Depending on the type of model, quench, and quantity monitored, the decay could be exponential, power law, a combination of both, or something else. To account for this lack of knowledge, we decided to fit a variety of functions to the data, including exponentials with and without an offset, power laws with and without an offset, and others. Frequently several of these functions fitted the data with equal accuracy (adjusted mean-squared deviation), and in these cases we determined the error bar of the prethermalized magnetization $\tilde{m}$ from the standard deviation around the mean value of the predictions of these different fits when extrapolated to later times. It is this rather conservative approach to the fitting that accounts for the fairly large error bars in Figs.~\ref{f:mofh} and \ref{f:mofh2}, but it makes sure that the phase diagrams are not biased by (possibly unjustified) assumptions about the functional form of the decay.

\bibliography{MK_Ref}
%----------------------------------------------
%\begin{thebibliography}{99}
%*********************************************************************************************
%\begin{thebibliography}{99}
%\end{thebibliography}
%--------------------------------------------------------
\end{document}